\documentclass[%
superscriptaddress,
preprint,
nofootinbib,
 amsmath,amssymb,
floatfix, %this somehow fixes some float issues. 
]{revtex4-1}

\usepackage{amsmath}
\usepackage{latexsym}
\usepackage{amssymb}
\usepackage{bbm}
\usepackage{amsthm}

\usepackage{graphicx,epstopdf,subfigure,color}
\usepackage{hyperref}
\usepackage{setspace}
\usepackage{natbib}
\usepackage{csquotes} % used for quotes \begin{displayquote}
\definecolor{Red}{rgb}{1,0,0}

\usepackage{qcircuit}
\usepackage[normalem]{ulem}

\def\bra#1{\mathinner{\langle{#1}|}}
\def\ket#1{\mathinner{|{#1}\rangle}}

\def\Bra#1{\left\langle#1\right|}

\def\BraVert{\egroup\,\mid\,\bgroup}

\def\Hil{\mathcal{H}}
\def\openone{\mathbbm{1}}
\def\idc{\mathfrak{1}}
\def\S{{\mathcal{S}}}
\def\M{{\mathcal{M}}}
\def\Me{{\mathfrak{M}}}
\def\R{{\mathfrak{R}}}
\def\D{{D}}
\def\BA{{\mathfrak{B}}}
\def\Un{{\mathfrak{U}}}
\def\In{{G}}
\def\SM{{_\mathcal{SM}}}
\def\En{{\mathcal{E}}}
\def\Ob{{\mathcal{O}}}

\def\SEO{{\mathcal{SEO}}}
\def\ASO{{\mathcal{ASO}}}
\def\A{{\mathcal{A}}}
\def\meas{{\Me}\left(U,\chi_{\En\Ob}\right)}

\def \quesurement {sensation} % terms that we might want to change
\def \quesurements {sensations}
\def \Quesurement {Sensation}
\def \Quesurements {Sensations}

\def \sensor {sensor}
\def \sensors {sensors}

\newtheorem{definition}{Definition}

\newtheorem{theorem}{Theorem}

\DeclareMathOperator{\tr}{tr}

\begin{document}

\title{
    Do qubits dream of entangled sheep? \\
    Quantum measurement without classical output
}

\author{Noah Lupu-Gladstein}
\email{nlupugla@physics.utoronto.ca}
\affiliation{Department of Physics and Center for Quantum Information and Quantum Control, University of Toronto, 60 St George St, Toronto, Ontario, M5S 1A7, Canada}
\author{Aharon Brodutch}
\email{brodutch@physics.utoronto.ca}
\affiliation{Department of Physics and Center for Quantum Information and Quantum Control, University of Toronto, 60 St George St,
Toronto, Ontario, M5S 1A7, Canada}
\affiliation{IonQ, Inc, College Park, MD, USA}
\author{Hugo Ferretti}
\affiliation{Department of Physics and Center for Quantum Information and Quantum Control, University of Toronto, 60 St George St, Toronto, Ontario, M5S 1A7, Canada}
\author{Weng-Kian Tham}
\affiliation{Department of Physics and Center for Quantum Information and Quantum Control, University of Toronto, 60 St George St, Toronto, Ontario, M5S 1A7, Canada}
\author{Arthur Ou Teen Pang}
\affiliation{Department of Physics and Center for Quantum Information and Quantum Control, University of Toronto, 60 St George St, Toronto, Ontario, M5S 1A7, Canada}
\author{Kent Bonsma-Fisher}
\email{kent.bonsma-fisher@nrc-cnrc.gc.ca}
\affiliation{Department of Physics and Center for Quantum Information and Quantum Control, University of Toronto, 60 St George St, Toronto, Ontario, M5S 1A7, Canada}
\affiliation{National Research Council of Canada, 100 Sussex Dr, Ottawa, Ontario, K1A 0R6, Canada}
\author{Aephraim M. Steinberg}
\email{steinberg@physics.utoronto.ca}
\affiliation{Department of Physics and Center for Quantum Information and Quantum Control, University of Toronto, 60 St George St, Toronto, Ontario, M5S 1A7, Canada}
\affiliation{Canadian Institute for Advanced Research, Toronto, Ontario, M5G 1M1, Canada}

\date{\today}

\begin{abstract}
Quantum mechanics is usually formulated with an implicit assumption that agents who can observe and interact with the world are external to it and have a classical memory. However, there is no accepted way to define the quantum-classical cut and no \emph{a priori} reason to rule out fully quantum agents with coherent quantum memories.  In this work, we introduce an entirely quantum notion of measurement, called a \emph{\quesurement{}}, to account for quantum agents that experience the world through quantum \sensors{}. \Quesurements{} eschew probabilities and instead describe a deterministic flow of quantum information. We quantify the information gain and disturbance of a \quesurement{} using concepts from quantum information theory and find that \quesurements{} always disturb at least as much as they inform. Viewing measurements as \quesurements{} could lead to a new understanding of quantum theory in general and to new results in the context of  quantum networks. 
\end{abstract}
%%%%%%%%%%%%%%%%%%%%%%%%%%%%%%%%%%%%%%%%%%%% END REMOVE %%%%%%%%%%%%%%%%%%%%%%%%%%%%%%%%%%%%%%%%%%%%%%%%%%%%%%%%%%%

\maketitle

\section{Introduction}

\begin{displayquote}
\it It is decisive to recognize that, however far the phenomena transcend the scope of classical physical explanation, the account of all evidence must be expressed in classical terms. \\ \begin{flushright}Niels Bohr \cite{BohrWheelerZurek}\end{flushright} 
\end{displayquote}

Measurement has been central to quantum mechanics since its inception. Historically, measurements were formalized with the notion of observables: Hermitian operators \cite{vonNeumann,VanHove1958} that represent quantities like position, momentum, and spin. This formulation of measurement has since been generalized several times \cite{KirkpatrickLeipzig, Davies1976, Wisemanbook, NielsenandChuang}, but none of these advances have questioned the basic notion that a measurement yields a classical outcome that can be copied. To account for classical outcomes, quantum mechanical models of measurement invoke an external, classical observer.

\begin{figure}
    \centering
    \includegraphics[width=0.8\textwidth]{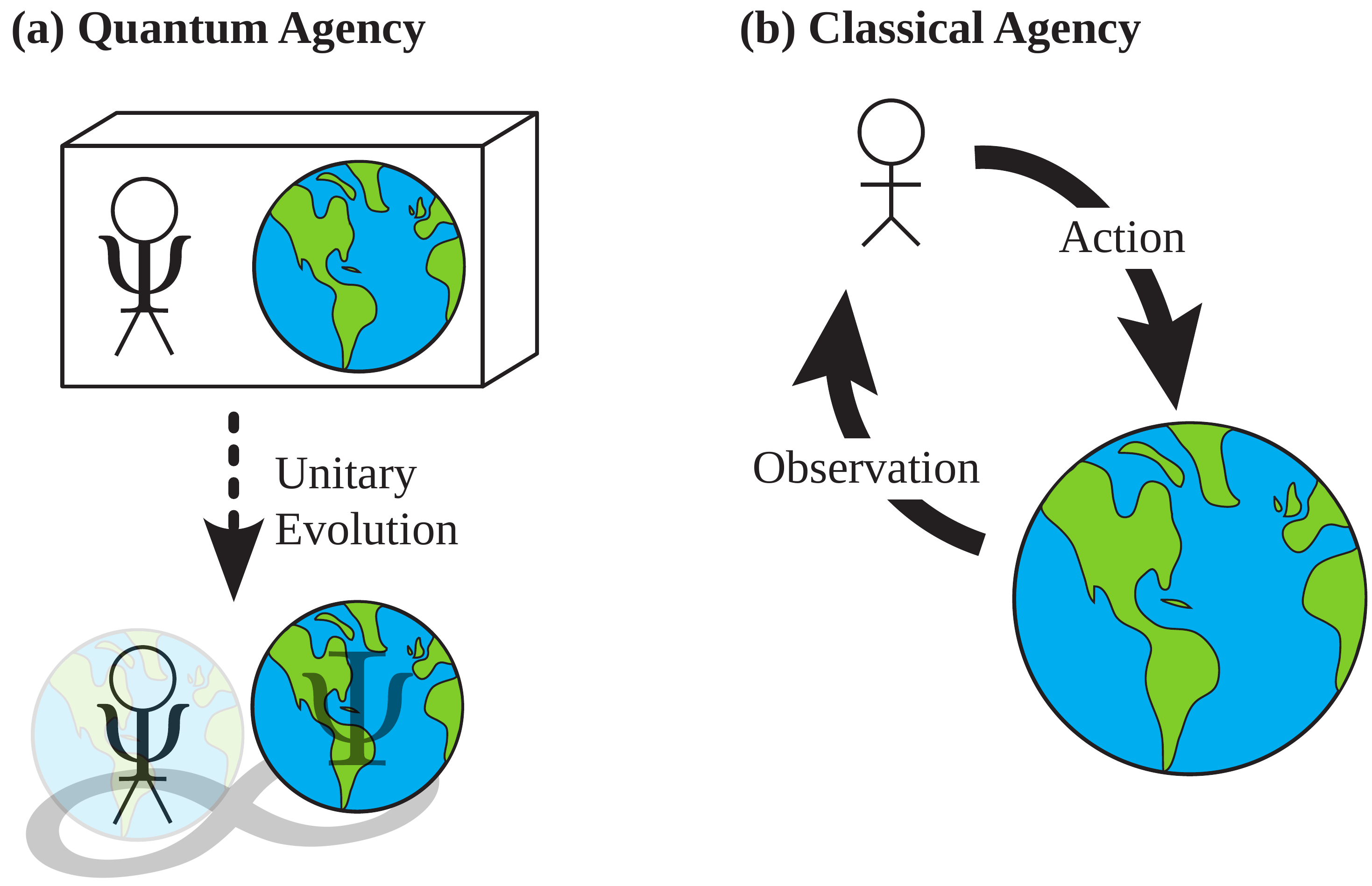}
   \caption{{\bf Agency with internal (\emph{quantum}) agents compared to external  (\emph{classical}) agents}. External agents can observe the world or act on it  (note that these these distinct operations can be inextricably connected through back-action). Quantum agents have quantum memories which are part of the quantum world. In the quantum scenario there is no clear distinction between an observation of the world and an action on the world.}
    \label{fig:agents}
\end{figure}

As quantum technology improves, we draw closer to an age where quantum machines make complex decisions using quantum computations that run on quantum data gathered with quantum sensors. These systems would be quantum agents in the sense that they autonomously sense and act upon the quantum world \cite{Dunjko2016, Kewming2021}. When a quantum agent measures its surroundings, is it really natural to view the results of that measurement as merely classical? Should the agent be viewed as external to the quantum world, or an active participant within it (see Fig. \ref{fig:agents})?

Our aim is to model quantum measurements as a quantum agent might: free of classical outcomes and external observers. A fully quantum paradigm will facilitate the development of new experiments, measurement techniques, communication protocols, and computational algorithms that do not fit easily in a classical framework. A complete description of quantum agents would not just model their observations, but also their decisions and actions. In this work, we focus solely on observation and leave the matter of decisions and actions for further study. We assume that a quantum agent has some ability to decide when and how to observe a given system. The mechanics behind this decision process are beyond the scope of this work, seeing as modeling how a classical agent makes decisions is already a difficult problem. We take for granted that a quantum agent can distinguish between its external environment and ``itself'', leaving the question of how a quantum agent even develops a sense of ``self'' for future research. Although this work does not paint a complete picture of quantum agency, its ideas have already inspired a forthcoming experiment \cite{pang2024agency}. 

Agents, both classical and quantum, experience the world through \sensors{}. We define a \sensor{} as a system with a memory that can interact with another system and store information from that system in its memory. We compare two types of  relationships between an observer and a \sensor{}:
\begin{itemize}
    \item {\bf Classical} - The \sensor{} is separate from the observer and the \sensor{}'s memory decoheres faster than the observer can process it
    \item {\bf Quantum} - The \sensor{} is a part of the observer and the \sensor{}'s memory remains coherent throughout being processed by the observer
\end{itemize}
For a given observer, we classify a \sensor{} as classical or quantum accordingly.
A quantum \sensor{} can take in incoherent information like a classical \sensor{}, but it can also take in and store quantum information about a quantum system.

As humans, we are naturally equipped only with classical \sensors{} and the standard notion of measurement (see Fig. \ref{fig:vNmeasurement}) reflects that. Traditionally, a measurement involves a system and an observable property of that system. The measurement is completed when an observer obtains a \emph{result}, i.e. a \emph{classical} record which can be copied and shared. Von Neumann developed a pragmatic quantum measurement model that is satisfactory for predicting our classical experience, yet invokes an observing agent external to the system \cite{vonNeumann}. There have been many attempts to adapt von Neumann's model and keep the observer inside the system (notably collapse theories and the approaches following Everett and Bohm \cite{Bohm1952,Stanford_phil_bohm_article,Everett57} ), but all cling to the central role of \emph{observables}.

We adopt a radical view of measurement: a measurement is any interaction between two systems in which the final state of one system depends on the initial state of the other. To avoid confusion with the historically laden term ``measurement'', we call such an interaction a \emph{\quesurement{}}. The result of a \quesurement{} is a quantum state, not a classical label. Completing a \quesurement{} does not require an external agent.

Any traditional measurement can be cast as a \quesurement{} followed by a transition to a classical value (the \quesurement{} is the so-called pre-measurement\cite{Busch1996}), but there are \quesurements{} that transcend traditional measurements. For example, consider a \sensor{} that swaps a state in its memory with a state in the environment. This interaction does not result in a distinct set of classical outcomes, so it cannot be associated with an observable or positive operator valued measure (POVM). It is nevertheless a \quesurement{} because the final state of the \sensor{} depends on the initial state of the environment.

We explore how an agent that uses \quesurements{} might view the world. We do not attempt to describe the inner workings of a quantum agent, nor attempt to accommodate the way humans with classical memories observe the world through events and probabilities. Instead, our aim is to describe information flow in a deterministic theory with as few assumptions as possible. In this sense our work is also very different from approaches that put the observer at the center (e.g \cite{Muller2020,Fuchs2014} ) which tend to begin with probabilities as a primitive. Our results give a consistent way of understanding information in a deterministic (collapse-free) interpretation of quantum mechanics as a complete theory in the sense that all physical objects (including observers) and all dynamical processes (including \quesurement{}) are described by the same rules.

Our work takes a world-view which is similar to the Everettian  (many-worlds, relative-state) interpretation, i.e., starting from Sec. \ref{sec:quantumObs} we will consider the \sensor{} as part of the quantum world and assume a unitary theory. However, our work is not an interpretation of quantum mechanics. It is instead the beginning of a framework for studying quantum agents.

 We begin in the next section with a description of the traditional --- classical \sensor{} --- approach to measurement.  In Sec. \ref{sec:quantumObs} we define \quesurements{} with quantum \sensors{}. We introduce the \emph{result channel} as the extension of the positive operator-valued measure (POVM) into the deterministic quantum regime, and  give two examples of \quesurements{}: the von Neumann \quesurement{}, and the swap \quesurement{} (Sec. \ref{sec:swap}). We then show how such swap \quesurements{} can be implemented in practice (Sec. \ref{sec:physicalmodel}). In Sec. \ref{sec:informationgain} we develop a method to quantify information gain, and disturbance. Using these tools, we compare the von Neumann and swap \quesurements{}. Implications of quantum agency are discussed in Sec. \ref{sec:agency} where we provide examples involving multiple agents sharing information, emphasising one of the limitations of the swap \quesurement{}, and mention the relation to quantum computing (Sec. \ref{sec:qcomputers}).  Our main conclusion (Sec. \ref{sec:conclusions}) is that a complete quantum treatment of agency which is free from anthropocentric pre-conceptions can lead to new results.  It is conceivable that by ignoring the possibility of quantum agents with quantum \sensors{} we are in danger of making an oversight similar to the one made by the founders of modern computing and information theory (many of whom had in-depth knowledge of quantum mechanics) who missed or ignored the possibility of quantum information processing.

\section{External observers: Observables and POVMs} \label{sec:externalobs}

\begin{figure}
    \centering
    \includegraphics[width=0.8\textwidth]{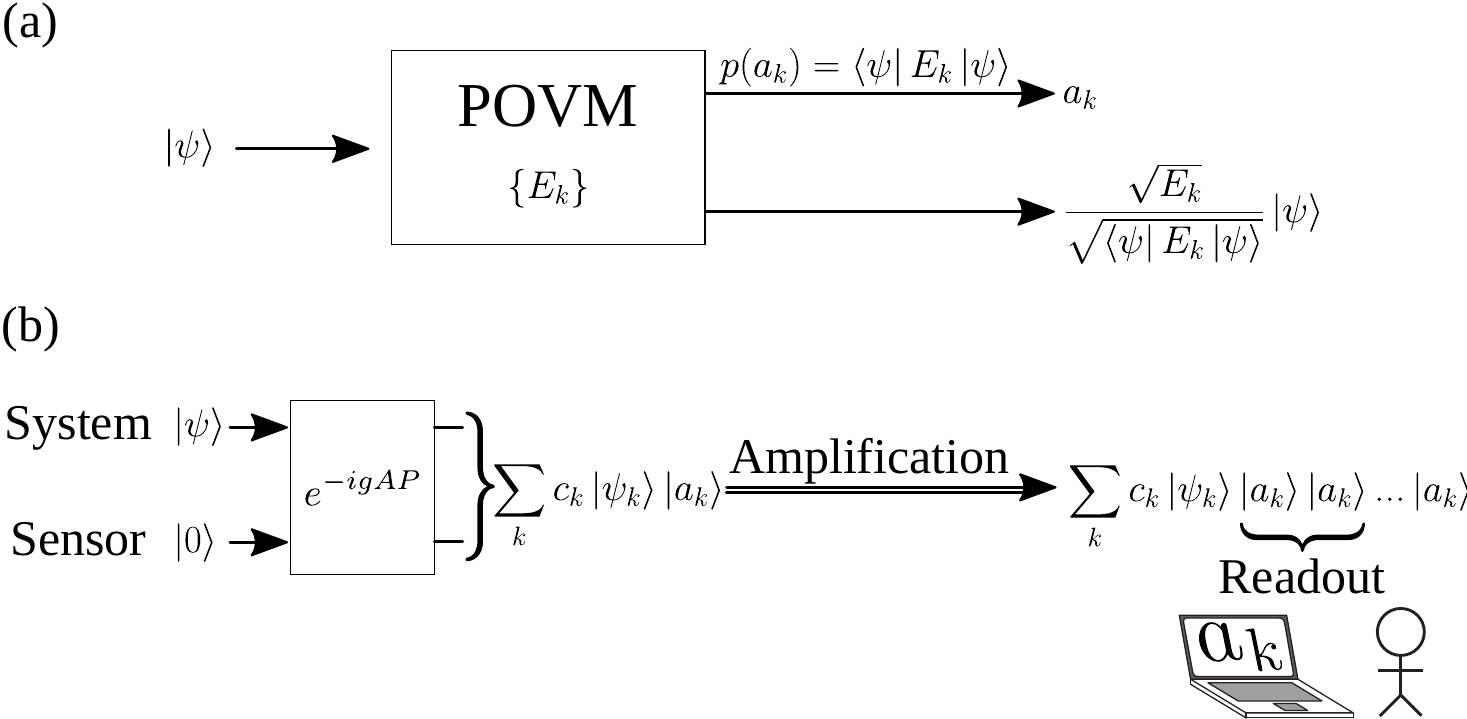}
    \caption{{\bf Quantum measurements with external agents.} (a) The traditional approach to  measurement  characterised by the POVM elements $\{E_k\}$. The probability for an outcome $a_k$ is $p(a_j)=\tr(E_k\rho)$ and the corresponding minimally disturbing \cite{Wisemanbook} state transformation is $\ket{\psi}\rightarrow \frac{\sqrt{E_k}\ket{\psi}}{\sqrt{\bra{\psi}E_k\ket{\psi}}}$. (b) The von Neumann scheme for a measurement of a non-degenerate observable $A$ provides a more detailed description than the textbook approach. It includes a quantum measurement device $\M$ and an amplification process whereby the information is copied onto multiple registers. The external observer reads out the state of some of these registers and records a classical result $a_k$. The cut between the external agent and the other subsystems has no observable consequences. 
     \label{fig:vNmeasurement} } 
\end{figure}

John von Neumann first described the measurement of an observable $A$ in terms of three separate subsystems:
\begin{enumerate}
    \item A system to be measured with an associated Hilbert space $\Hil_\S$  on which the observable $A$ is an Hermitian operator
    \item A \sensor{} with Hilbert space $\Hil_\M$ on which there are two canonically conjugate operators: the pointer operator $P_\M$ and its canonical conjugate $Q_\M$
    \item An external agent with no mathematical representation
\end{enumerate} 
The measurement process according to von Neumann's scheme can be broken into two stages: interaction and readout. In the interaction stage a Hamiltonian of the type  
 \begin{equation}\label{eq:vonN}
    H_i \propto A_\S \otimes Q_\M
 \end{equation} 
is switched on to couple the system and \sensor{}. The result can then be amplified (see Fig. \ref{fig:vNmeasurement} b). At the readout stage, an external agent records the state of the \sensor{}.  Prior to the mathematical derivation, von Neumann (invoking Bohr \cite[footnote 207]{vonNeumann}) argues that the agent being external is not unique to the quantum regime, and that the precise cut between the agent and the \sensor{} is arbitrary even in the classical case. The aim of the derivation, which would later be called the von Neumann measurement scheme, was to regain the classical intuition that measurements are independent of the specific choice of the cut between the \sensor{} and the agent. This \emph{motility of the cut} is demonstrated by showing that the agent and the \sensor{} can be treated as a single composite system which can be cut into subsystems arbitrarily without modifying the outcome probabilities or the state update rules.

To make a comparison between \quesurements{} and traditional measurements, we list a number of features that arise from the von Neumann scheme:

\begin{itemize}
\item Result - After the interaction, the \sensor{}'s memory is in a state that generally depends on the initial system. 
\item Broadcastability - The result can be copied and broadcast to others.
\item Constrained back-action - A second \sensor{} interacting identically with the same system yields the same result up to statistical uncertainty.
\item Motility of the cut - There is no accepted scientific theory that separates the \sensor{} from the external observer.  
\end{itemize}

The first feature is essential in the definition of a measurement \cite{Everett57}. The next two points 
play a key role in communication since they allow multiple agents to communicate the results of their measurements by making copies and broadcasting them. Such communication engenders objectivity in the sense that different agents can agree on some specific property by making individual measurements of the same system and comparing their results \cite{Brand_o_2015,Everett57,Ollivier_2004,Fields_2013}.  
Historically, these or some subset of these features are taken as the defining properties of measurements  (e.g. \cite{Deutsch_2015}). Our aim in this work is to explore the consequences of keeping only the requirement that measurements have results that depend on the initial system.

\section{The quantum observer}\label{sec:quantumObs}

The features of von Neumann measurements match our classical experience, but do they also capture everything a quantum observer can experience? To start answering this question, we define \quesurements{}, the more general class of interactions enabled by access to a quantum \sensor{}.

\begin{figure}
    \centering
    \includegraphics[width=0.8\textwidth]{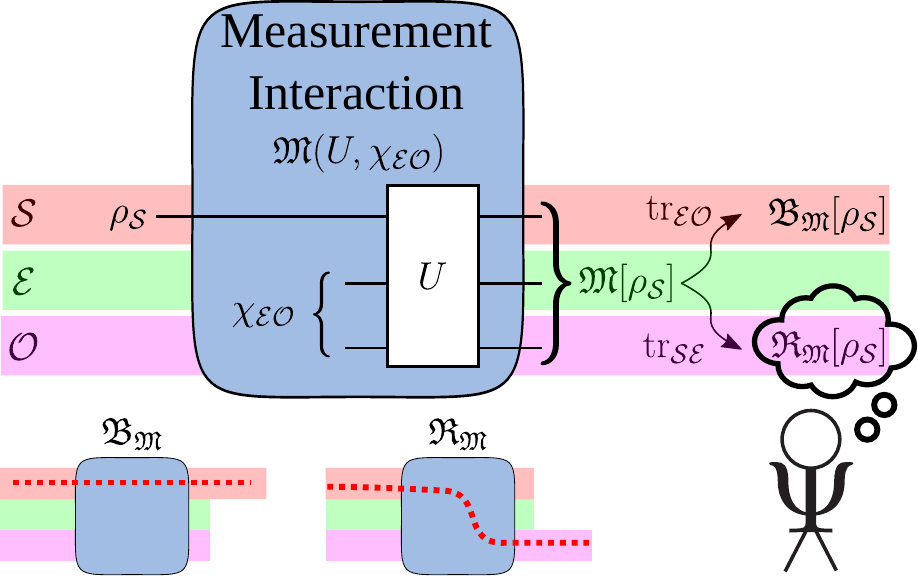}
    
    \caption{{\bf Quantum measurements with a quantum agent:} A measurement interaction $\Me$ (blue box) fills an observer's memory $\Ob$ with information about a system $\S$ in the presence of an environment $\En$. The interaction is defined in terms of two objects: a unitary $U$ (which couples the system $\S$, environment $\En$ and the observer's memory $\Ob$), and the initial $\En\Ob$ state $\chi_{\En\Ob}$.  The interaction takes an initial  system state $\rho_\S$ to a final $\S\Ob$ state $\Me[\rho_\S]$. The result $\R_\Me[\rho_\S]$ is the reduced state encoded in the observer's memory. The interaction induces the result channel $\R_\Me$ (dashed red on bottom right) from system states to observer memory states and the back-action $\BA_{\Me}$ (dashed red on bottom left), a channel from system states to disturbed system states. The entire process is deterministic. The interaction is considered a \quesurement{} (according to Def. \ref{def:meas}) as long as $\R_\Me[\rho_\S]$ is not a constant function of $\rho_\S$, i.e., $\Me$ is a \quesurement{} of $\S$ whenever the result depends on the state of $\S$.  
    }
    \label{fig:quantummeasurement}
\end{figure}

\subsection{Defining \quesurement{}}

As before, the system $\S$ to be observed is associated with the Hilbert space $\Hil_\S$. The \sensor{} is a part of an observer $\Ob$ and interacts with the system. 
In contrast to traditional measurement, we consider the observer itself a quantum system. We associate the memory in the observer's \sensor{} with the Hilbert space $\Hil_\Ob$. 
The \quesurement{} is a procedure which includes a unitary operation $U$ on $\S$, $\Ob$, and possibly other subsystems. We call these other subsystems the environment $\En$, with Hilbert space $\Hil_\En$, because it is conceptually separate from the measured system and the observer. Despite the terminology, the environment need not be noisy and could simply represent an additional \sensor{} (see section \ref{sec:vNexample}). 

For simplicity, we assume that  free evolution can be ignored on the time scales of the \quesurement{}\footnote{Note that in practice the free evolution can lead to significant corrections, see for example Sec. \ref{sec:photondetector} below.}. To  keep the notation simple, we will also use no subscripts when describing an operator  on the joint $\SEO$ Hilbert space $\Hil_\S\otimes\Hil_\En\otimes\Hil_\Ob$. Furthermore, we will use the term channel to refer to completely-positive trace-preserving maps. We denote such channels  $\mathfrak{C}:\mathcal{X}\rightarrow\mathcal{Y}$ when they take states  on $\Hil_\mathcal{X}$ to states on $\Hil_\mathcal{Y}$.

A \quesurement{} is driven by an interaction between the system, observer, and environment. The interaction is defined by a unitary $U$ and the initial composite state $\chi_{\En\Ob}$ of the environment $\En$ and the observer $\Ob$. Mathematically, the interaction is  a quantum channel $\meas:\S\rightarrow\S\Ob$ acting on an arbitrary system state $\rho_\S$ \begin{equation}
\meas[\rho_\S] = \tr_{\En} \left [U \rho_\S \otimes \chi_{\En\Ob} U^\dagger \right ]
\end{equation} 
(See Fig. \ref{fig:quantummeasurement} a).
To simplify our notation we will henceforth drop the arguments in $\meas$. 

For any interaction $\Me:\S\rightarrow\S\Ob$, we also define a \emph{result channel}  $\R_\Me : \S \rightarrow \Ob$ that maps the state $\rho_\S$ onto a state in $\Hil_\Ob$ called the result,   
%\R\Me(ρ§)=\tr§\En[Uρ§⊗χ\En\ObU†]\R_\Me(\rho_\S)= \tr_{\S\En}\left[U \rho_\S \otimes \chi_{\En\Ob} U^\dagger\right] 
\begin{equation}\label{eq:defphi}
\R_\Me[\rho_\S]= \tr_{\S}\left[ \Me\left[\rho_\S\right] \right ].
\end{equation}
This result channel can be seen as the analogue of the POVM,  but whereas the POVM maps system states to probabilities, the result channel $\R_\Me$ maps system states to observer states. Similarly, we introduce the \emph{back-action} 
\begin{equation}\label{eq:lambda}
    \BA_\Me[\rho_\S]=\tr_{\Ob}\left[\Me[\rho_\S]\right]
\end{equation} 
which is a channel that describes how the interaction modifies $\S$. This channel is the standard channel associated with measurement back-action (as defined in \cite{Wisemanbook} for example).
In a theory with external agents we would say that a channel is a measurement if the associated POVM elements are not all proportional to the identity, similarly the result channel plays the central role in  the following definition of  a measurement.

\begin{definition}[\Quesurement{}, result]\label{def:meas}
An interaction $\Me:\S\rightarrow\S\Ob$ with an associated result channel $\R_\Me:\S\rightarrow\Ob$ is a \emph{\quesurement{}} (of $\S$ by $\Ob$) if and only if there are two system states $\rho_\S,\sigma_\S$ such that $\R_\Me[\rho_\S]\ne \R_\Me[\sigma_\S]$. The \emph{result} is the quantum state $\R_\Me[\rho_\S]$. 
\end{definition}

Note that since the channel is linear, it is sufficient to consider pure states. That is, a channel $\Me:\S\rightarrow\S\Ob$ with an associated result channel $\R_\Me:\S\rightarrow\Ob$ is a \quesurement{}  when there are two pure states $\rho_\S, \sigma_\S$ such that $\R_\Me[\rho_\S]\ne\R_\Me[\sigma_\S]$.

The asymmetry in our definition of \quesurements{} as channels from $\S$ to $\S\Ob$ (as opposed to $\S\Ob$ to $\S\Ob$) may seem like a ``bug'', but it is actually a crucial feature. It provides a formal lever to distinguish the observer from the observed.  If two quantum agents interact, they will each experience a different \quesurement{} because the two agents will define themselves as the observer and the other as the system. A coupling between two agents $\mathcal{A}$ and $\mathcal{B}$ establishes an \emph{objective} channel from $\mathcal{A}\mathcal{B}$ to $\mathcal{A}\mathcal{B}$, but the resulting \quesurement{} is \emph{subjective} in that it depends on whether $\mathcal{A}$ is seen as the observer or the system.

We will soon explore concrete examples of specific \quesurements{}, but first we will give two examples of interactions that are not \quesurements{}. An interaction driven by any unitary of the form $U_{\SEO}=U_{\S\En} \otimes U_{\Ob}$, is not a \quesurement{}. Second, consider an interaction of the form $U_{\SEO}=e^{i X_\S\otimes Y_{\En\Ob}}\neq \idc$. The interaction is not a \quesurement{} in the special case where  $\chi_{\En\Ob}$ happens to commute with $Y_{\En\Ob}$, although it is a \quesurement{} otherwise.

\subsubsection{Example 1: the von Neumann measurement} \label{sec:vNexample} 

\begin{figure}
    \centering
    \includegraphics[width=0.8\textwidth]{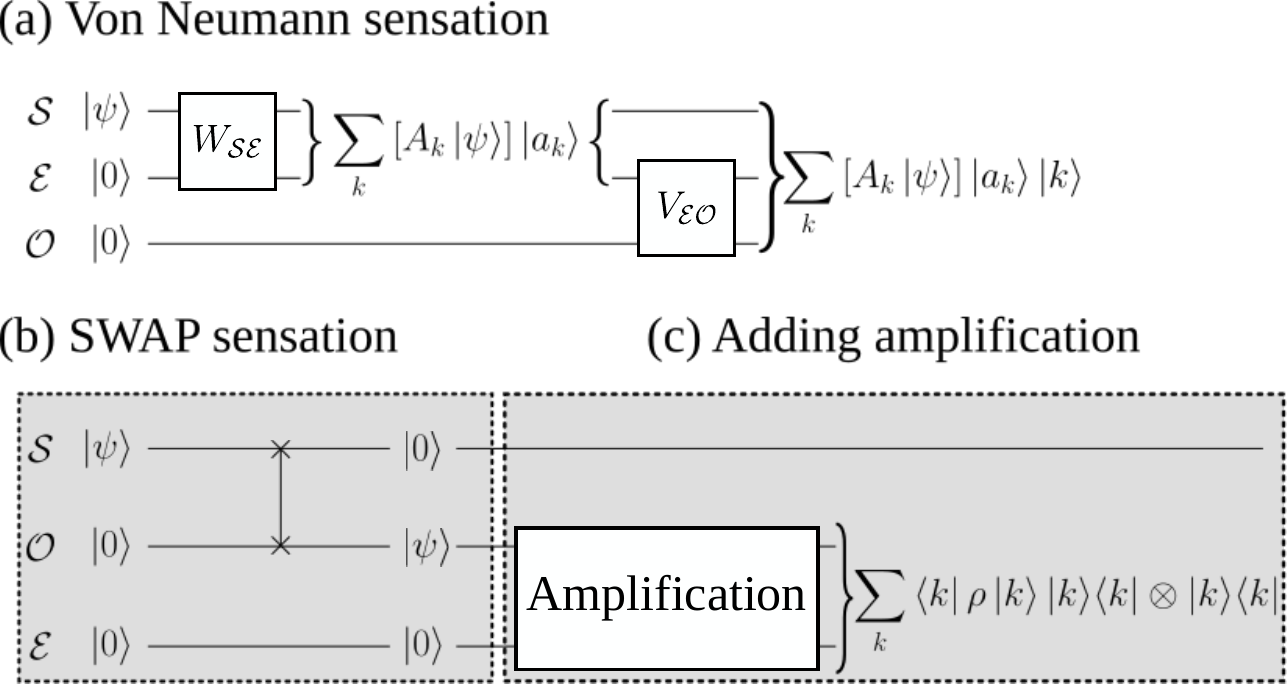}
    \caption{{\bf Examples of measurements with a quantum agent:} (a) The circuit diagram for a von Neumann measurement performed in two steps (Eq. \ref{eq:vnmeasurement}). The first, step $W_{\S\En}$ couples between the environment (or measuring device) and the system. The second, $V_{\En\Ob}$, couples between the environment and the observer.  The result is a quantum state encoded in a preferred basis $\{\ket{k}\}_k$ which can be copied (amplified) and broadcast. Compare this to  von Neumann's original approach  (Fig. \ref{fig:vNmeasurement}), where the observer is external to quantum dynamics and the result is a classical label.   (b) The circuit diagram for a  swap measurement. The observer learns everything about the system ($\ket{\psi}$ is now encoded in the memory), but the disturbance is maximal (the system retains no trace of its original state). There is no preferred basis and so the state cannot be copied and broadcast.  (c) Amplifying in a specific basis  $\{\ket{k}\}_k$ causes decoherence. The result can be copied and broadcast but phase information is lost. 
  }
    \label{fig:examples}
\end{figure}

In the von Neumann measurement, the environment is a classical \sensor{}. To measure an observable  $A=\sum_k a_k A_k$ with eigenspace projectors $A_k$, we start with an initial  product state written as $\ket{\psi}\ket{0}\ket{0}$, where $\ket{\psi}$ is an arbitrary $\S$ state and $\ket{0}\ket{0}$ is a fixed initial $\En\Ob$ state. We then generate an interaction $W_{\S\En}$ between the  environment and the system  and follow with an interaction  $V_{\En\Ob}$ between the system and the observer, so  $U=V_{\En\Ob}W_{\S\En}$. These interactions and the initial $\En\Ob$ states are chosen so that  
\begin{align}\label{eq:vnmeasurement}
\ket{\psi}\ket{0}\ket{0}\overset{W_{\S\En}}{\longrightarrow} \sum_k [A_k\ket{\psi}]\ket{a_k}\ket{0}\overset{V_{\En\Ob}}{\longrightarrow}\sum_k[A_k\ket{\psi}]\ket{a_k}\ket{k}.
\end{align} 
where the  environment states $\ket{a_k}$ are orthogonal to each other, as are the memory states $\ket{k}$ (we assume that the dimensions of $\Hil_{\En}$ and $\Hil_{\Ob}$ are large enough).  Each of these interactions can be generated by a Hamiltonian in the form of Eq. \eqref{eq:vonN}. 

The result of the interaction is the final memory state $\sum_k \bra{\psi}A_k\ket{\psi}\ket{k}\bra{k}$.  The observer's information about the system is captured by the correlations between $\S$ and $\Ob$. The result depends on the initial system state in the sense that each of the orthogonal \sensor{} states `points' at a corresponding system state with weights determined by the initial system state.

\subsubsection{Example 2: The swap \quesurement{}} \label{sec:swap}

We now go to the extreme situation where the agent learns  everything about the quantum state of the system by ``swapping'' the latter, lock, stock, and barrel with her quantum \sensor{}'s memory register. We refer to this operation as the \emph{swap \quesurement{}}. For simplicity, we choose $\Ob$ to be the same dimension as  $\S$ so  $U$ can be the standard SWAP gate \cite{NielsenandChuang} between $\S$ and $\Ob$. For  an initial  $\SEO$ state  $\ket{\psi}\ket{0}\ket{0}$, the SWAP induces the transformation $\ket{\psi}\ket{0}\ket{0}\rightarrow \ket{0}\ket{0}\ket{\psi}$ (see Fig. \ref{fig:examples} b) which puts the state $\ket{\psi}$ in the observer's \sensor{}. Following this measurement, the observer' \sensor{} contains everything about the state of $\S$ before the interaction, at the cost of a very strong back-action. The \quesurement{}'s result cannot be shared across multiple \sensors{} and there is no system observable associated with it. The agent senses \emph{the state} $\ket{\psi}$, rather than a specific property of $\ket{\psi}$. 

The back-action of a swap \quesurement{} is so strong, one may wonder whether it is even appropriate to call $\S$ the system once its contents have been entirely swapped with a part of the observer. In practice, $\S$ is often a small piece of a much larger system. For example, if an observer uses a swap \quesurement{} to store a photon in a quantum memory (explored further in Sec. \ref{sec:photondetector}), she only senses one small portion of the entire electromagnetic field. Whatever state was swapped out from her \sensor{}'s memory would start to participate in the dynamics of electromagnetism, so there is a physical sense in which $\S$ is still the system, even after its contents have been completely disturbed.

\subsubsection{Example 3: The decohered swap} \label{sec:decoheredSwap}

A  more realistic scenario  involves a memory which is open to the environment. Analysis in this case depends  on the precise dynamics of the coupling to the environment.  One possibility that  allows us to recapture the classically intuitive broadcasting feature is dephasing of $\Ob$ in some preferred basis $\{\ket{k}\}_k$ through interaction with the environment (see Fig \ref{fig:examples} c). The full transformation for the $\S\Ob$ subsystems would then be
\begin{equation}\label{eq:dephasedswap}
\rho_\S\otimes\tau_\Ob\rightarrow\tau_\S\otimes\rho_\Ob\rightarrow\tau_\S\otimes\sum_k\bra{k}\rho\ket{k}\ket{k}\bra{k}_\Ob
\end{equation}
The observer's state  is then similar to the one in the von Neumann scheme above (Eq. \ref{eq:vnmeasurement}) with equivalence in the case $A_k=\ket{k}\bra{k}$, however there are no longer any correlations with $\S$. Correlations with $\En$ would be similar to those of Eq. \eqref{eq:vnmeasurement} and in principle $\En$ could include many copies of the result.  As we will show below, the swap-and-decohere process is similar to the behaviour of a photon detector. The main feature we want to highlight for now is that once decoherence kicks in, the \quesurement{} can be associated with an observable like a traditional measurement. The main features of all three \quesurements{} discussed above are presented in Table \ref{tab:measurements} using  terminology which we  develop further in section \ref{sec:informationgain}. 

\subsection{Physical models for a swap \quesurement{}}\label{sec:physicalmodel}

The swap \quesurement{} is an extreme example of a new capability granted by quantum \sensors{} with ideal quantum memories. We now show that this \quesurement{} is not only physically feasible, but that the dephased swap is also fairly commonplace. 

\subsubsection{Photon detectors} \label{sec:photondetector}

Photoelectric devices that measure light intensity are among the most common ways to sense the world. These devices typically work by swapping the state of the incoming light field with the state of photoelectrons, which then interact with each other and rapidly decohere (see Appendix.~\ref{sec:photodetection}). This detection process is essentially the decohered swap measurement described above in Sec.~\ref{sec:swap}. The fact that one of the  most ubiquitous ways of observing the world does not follow the von Neumann scheme and moreover acts like a decohered SWAP operation (with all information in the field being lost) is significant. In particular any assertion (see \cite{Everett57} for example) that good observations should be repeatable by multiple observers must be assessed  with this in mind. While photodetection with rapid decoherence does not fit within the narrow model of von Neumann measurements, it is nevertheless a classical measurement in the sense that its outcomes can be modelled as classical labels rather than coherent quantum states. Only the number (as opposed to the relative phase) of photoelectrons detected is meaningful because the environment washes out any definite phase relation between photoelectron number states.

\subsubsection{Optical quantum memories}

\begin{figure}
\center{\includegraphics[width=0.8\linewidth]{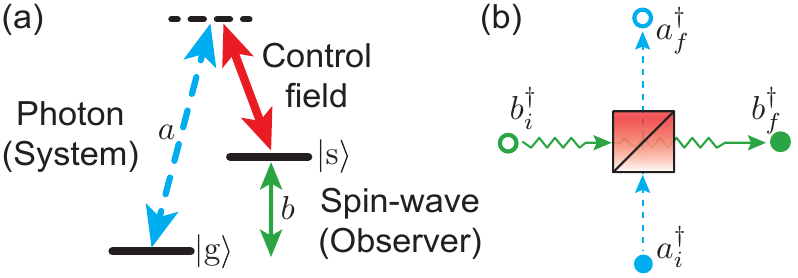}}
\caption{ (a) The 3-level system for an off-resonant Raman quantum memory. A flying photon mode $a$ is mapped, via a strong control field, to a stationary spin-wave excitation. The spin-wave mode $b$ is an excitation from the ground state  ($\ket{g}$)  to the  storage state ($\ket{s}$)  of the medium. 
(b) The storage process can be represented as a beamsplitter interaction between the optical and spin-wave states. With unit reflectivity, i.e., storage efficiency, number information is swapped between modes $a$ and $b$:  $b^\dagger_i \to a^\dagger_f$ and $a^\dagger_i \to b^\dagger_f$.
}
\label{fig:memorybs}
\end{figure}

If the environment's action on a photodetector can be sufficiently tamed, it becomes more than a mere measurement. It becomes a quantum memory for an optical mode. Such a memory is not beholden to any particular optical observable. Its dynamics are described by \quesurements{} where the optical mode is the system to be measured, and the memory mode represents the \sensor{}. 
It has been demonstrated that photon storage in an off-resonant Raman quantum memory is equivalent to a beamsplitter interaction between a flying photon mode ($a^\dagger$) and a stationary spin-wave excitation ($b^\dagger$)  \cite{Reim2012,England2016} (see Fig. \ref{fig:memorybs}).
The `reflectivity' in this beamsplitter interaction is given by the storage efficiency of the memory which can, in theory, approach unity \cite{Nunn2007}.
This storage process acts like a SWAP  between the bosonic optical and spin-wave memory modes. Any superposition of photon number states in the optical mode becomes the same superposition of spin-wave excitations in the spin-wave mode and vice versa.

Perfect or near-perfect quantum memories are yet to be demonstrated, but it is reasonable to expect that such devices will exist in the not-too-distant future, most likely with some type of error correction mechanisms to increase coherence times. Some of these memories, in particular those directly connected to communication channels and sensors, will likely be optical memories with a mechanism that resembles a SWAP operation with an incoming light mode. A sufficiently advanced quantum computer might interact with the world mostly through swap \quesurements{} (see Sec.  \ref{sec:qcomputers}).

%%% FIG %%%

\begin{table}[ht]
\begin{center}
\begin{tabular}{ l | c c c}
  & von Neumann & SWAP & SWAP  \\
  &&&with decoherence\\
  \hline\\

 Motility of the cut & Yes & No & Yes\\  
 Broadcastable & Yes & No & Yes\\
 Repeatable   & Yes & No & No\\
 Reversibility & Requires  environment & Yes & Requires  environment \\
 Information gain $\In(\Me)$ &  Heisenberg  & Maximal & Heisenberg \\
 Disturbance $\D(\Me)$ & Heisenberg &  Maximal & Maximal
%\\ Information gain  I\A:\ObI_{\A:\Ob} &log2dlog_2 d &2log2d2log_2 d &log2dlog_2 d
\end{tabular}
\end{center}
\caption{Comparing three types of \quesurements{}: the von Neumann scheme, the swap, and the swap with decoherence (see Sec. \ref{sec:vNexample},\ref{sec:swap} for details). Motility of the cut refers to the possibility of identifying different cuts between the observer and the environment. The result is broadcastable when it can be copied and shared with other observers, and the measurement is repeatable if a second measurement of the same type will produce the same outcome. Reversibility refers to the resources required for reversing the operation so that the system will be restored to its original state (the resources are either access to $\S\Ob$ or access the entire $\SEO$).  Information gain and disturbance (Defined in Sec.  \ref{sec:quantifying}) are given in terms of two reference points defined in Sec. \ref{Sec:uncertainty}. For a system of dimension $d$, the Heisenberg limit is $\log d$ and maximal is $2 \log d$. The swap \quesurement{} obtains maximal information at the cost of maximal disturbance. The decohered swap performs as well or worse than the von Neumann scheme on all accounts and is essentially equivalent to a destructive von Neumann measurements like photodetection. 
}
\label{tab:measurements}
\end{table}

\section{Information Theory for Quantum Observers} \label{sec:informationgain}

The information gained from a measurement is usually quantified through some function of the probability distribution associated with the possible measurement results, viewed as classical labels. This approach cannot be applied as is to our deterministic framework that defines results as quantum states. In this section we present an intuitive way to think about information gain and disturbance that can be applied in the quantum-agent scenario. We start by introducing \emph{maximally informative}  and \emph{maximally disturbing} \quesurements{} in Sec. \ref{Sec:uncertainty} and point out that their non-informative and non-disturbing counterparts are not \quesurements{}. We then (Sec \ref{sec:quantifying}) introduce an additional agent  $\A$ who is initially entangled with $\S$ (see Fig \ref{sec:informationgain}) and show that correlations with this agent can be used to quantify information gain and disturbance. We calculate the information gain and disturbance for the von Neumann, swap, and decohered swap \quesurements{} (Sec. \ref{sec:examples}). We discover, and subsequently prove, that even the most exotic \quesurement{} can never inform more than it disturbs.

\begin{figure}
\center{\includegraphics[width=0.8\linewidth]{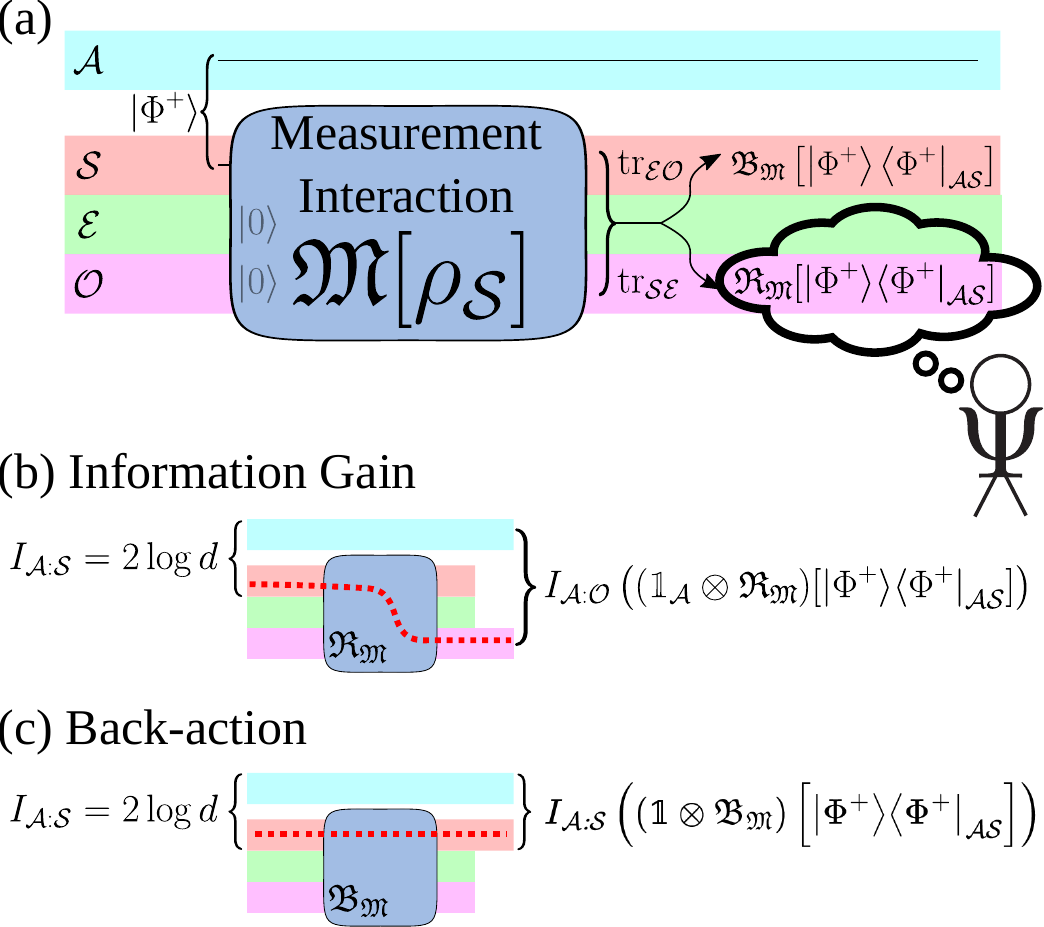}}
\caption{ {\bf Information gain and disturbance}: (a) A second agent $\A$  is initially entangled with $\S$ so that they are maximally correlated, $I_{\A:\S}= 2 \log d$. The observer $\Ob$ is initially uncorrelated with $\A$. (b) The information gain is defined as the change in mutual information between $\A$ and $\Ob$ after the \quesurement{} (Eq. \ref{eq:uncertainty}). (c) The disturbance is the change in mutual information between $\A$ and $\S$ (Eq. \ref{eq:backaction}).}   
\label{fig:informationg}
\end{figure}

\subsection{Maximal information gain and disturbance} \label{Sec:uncertainty}

A \quesurement{} $\Me$ is \emph{maximally informative} whenever the observer's final memory state and the details of $\Me$ are sufficient to produce at least one instance of a state identical to any arbitrary initial system state.
\begin{definition}[Maximally informative \quesurement{}] \label{def:maximallyinformative}
A \quesurement{} $\Me: \S \rightarrow \S\Ob$ with an associated result channel $\R_\Me: \S \rightarrow \Ob$ is called maximally informative when there exists a quantum channel $\R^{-1}_\Me: \Ob \rightarrow \S$ such that the composition of channels $\R^{-1}_\Me\circ \R_\Me=\idc$ where $\idc$ is the identity channel.
\end{definition}
It is tempting to demand that a maximally informative \quesurement{} enables the observer to produce not just one, but any number of states identical to a given input. While that requirement might be appropriate for a classical observer, quantum observers must abide by the no-cloning theorem \cite{wootters1982single, dieks1982communication}. A related, but distinct notion in standard quantum mechanics is an informationally complete measurement \cite{d2004informationally}. These measurements eventually allow an observer to produce many copies of a measured state, but only after measuring many identical instances of that state.

The opposite limit of a maximally informative \quesurement{} is an interaction in which the observer's final state is independent of the initial system state. Such a \emph{non-informative interaction} is in fact not a \quesurement{} under Def. \ref{def:meas}.
\begin{definition}[Non-informative interaction] \label{def:noninformative}
An interaction $\Me:\S \rightarrow \S\Ob$ with an associated result channel $\R_\Me: \S \rightarrow \Ob$ is called non-informative when $\R_\Me[\rho_\S] = \R_\Me[\sigma_\S]$ for any pair of states $\rho_\S, \sigma_\S$ in $\S$. 
\end{definition} 

A \quesurement{} is \emph{maximally disturbing} if the final system state is independent of the initial system state.
\begin{definition}[Maximally disturbing \quesurement{}]\label{def:maximallydisturbing}
A \quesurement{} $\Me:\S\rightarrow \S\Ob$ with an associated back-action $\BA_\Me: \S \rightarrow \S$ is called maximally disturbing when $\BA_\Me[\rho_\S] = \BA_\Me[\sigma_\S]$ for any pair of states $\rho_\S, \sigma_\S$ in $\S$. 
\end{definition} 

By contrast, a \emph{non-disturbing interaction} is one that allows the system to perfectly recover its initial state from its final state.
\begin{definition}[Non-disturbing interaction] \label{def:nondisturbing}
An interaction $\Me:\S\rightarrow \S\Ob$ with an associated back-action $\BA_\Me:\S \rightarrow \S$ is called non-disturbing when there exists a quantum channel $\BA^{-1}_\Me: \S \rightarrow \S$ such that  the composition of channels $\BA^{-1}_\Me\circ \BA_\Me=\idc$ where $\idc$ is the identity channel. 
\end{definition}

\subsection{Quantifying information gain and disturbance} \label{sec:quantifying}

We seek an information gain function $\In(\R_\Me)$ and a disturbance function $\D(\BA_\Me)$ to quantitatively describe the results and back-action of an interaction $\Me$. These functions must be compatible with the definitions given above for maximally informative, non-informative, maximally disturbing, and non-disturbing. Furthermore, we demand that no local operation on $\Ob$ after the interaction increases $\In(\R_\Me)$. Similarly, no local operation on $\S$ after the interaction may decrease $\D(\BA_\Me)$. These requirements are summarized below.
 \begin{itemize}
    \item $\In(\R_\Me)=0$ when $\Me$ is non-informative and strictly positive otherwise. 
    \item $\In(\R_\Me)$ maximizes for a given system $\S$ when $\Me$ is maximally informative.
    \item $\In(\R_\Me)$ is non-increasing under local operations on $\Ob$:  if  $\Me'=\mathfrak{C}_\Ob\circ\Me$ and $\mathfrak{C}_\Ob$ is a local channel on $\Ob$, then $\In(\Me')\le\In(\Me)$. 
 \end{itemize}

\begin{itemize}
    \item $\D(\BA_\Me)=0$ when $\Me$ is non-disturbing and strictly positive otherwise. 
     \item $\D(\BA_\Me)$ maximizes for a given system $\S$ when $\Me$ is maximally disturbing.
     \item $\D(\BA_\Me)$ is non-decreasing  under local operations on $\S$:  if  $\Me'=\mathfrak{C}_\Ob\circ\Me$ and $\mathfrak{C}_\Ob$ is a local channel on $\Ob$, then $\D(\Me')\ge\D(\Me)$. 
 \end{itemize}

To construct functions $\In(\R_\Me)$ and $\D(\BA_\Me)$ that respect the above properties, we will consider a second agent $\A$ who is maximally entangled with $\S$ before the measurement interaction. Maximally entangling $\S$ with $\A$ reflects the assumption that the observer has no prior knowledge about the system state $\rho_\S$. \footnote{It is possible to modify the approach by encoding any prior knowledge the observer has in the initial $\A\S$ state so that $\A\SEO$ would be the purification of the ensemble of initial states, but the details of such modifications are beyond the scope of this work.} For concreteness, we choose the entangled state to be $\ket{\Phi^+}_{\A\S}=\frac{1}{\sqrt{d}}\sum_{k = 1}^d \ket{k}_\A\ket{k}_\S$ where $d=\dim(H_\S)=\dim(H_\A)$ and $\{\ket{k}\}_{k = 1}^d$ is a complete, orthonormal basis. \footnote{For the sake of simplicity, we consider only finite dimensional Hilbert spaces.}  We will quantify the correlations between two subsystems $\mathcal{X}$ and $\mathcal{Y}$ via mutual information, defined as 
\begin{equation}
    I_{\mathcal{X}:\mathcal{Y}}(\tau_{\mathcal{X}\mathcal{Y}})=
    S(\tau_{\mathcal{X}}) + S(\tau_{\mathcal{Y}}) - S(\tau_{\mathcal{X} \mathcal{Y}})
\end{equation}
where $S(\tau)=-\tr\left[\tau\log\tau\right]$ is the von Neumann entropy of a state $\tau$.
Mutual information enjoys some properties \cite{Vedral_2002,NielsenandChuang,BrodutchModi} which are specifically useful for our purposes:

\begin{itemize}
    \item $I_{\mathcal{X}:\mathcal{Y}}(\tau_{\mathcal{X}\mathcal{Y}}) \geq 0$ with equality if and only if $\tau_{\mathcal{X}\mathcal{Y}}$ is a product state, i.e. $\tau_{\mathcal{X}\mathcal{Y}}=\tau_{\mathcal{X}}\otimes\tau_{\mathcal{Y}}$. 
    \item The mutual information reaches its maximal value $2 \log d$ when $\tau_{\mathcal{X}\mathcal{Y}}$ is a maximally entangled state. 
    \item Mutual information is non-increasing under local operations.
\end{itemize}

We can now quantify both the information gain and back-action by looking at the change in mutual information after the measurement interaction. We define information gain $\In(\R_\Me)$ as $I_{\A:\Ob}$ after the interaction $\Me$
\begin{equation} \label{eq:uncertainty}
    \In(\R_\Me) \equiv I_{\A:\Ob}\left( \left(\idc_\A \otimes \R_\Me \right) [\ket{\Phi^+}\bra{\Phi^+}_{\A\S}] \right).
\end{equation}
and the disturbance $\D(\BA_\Me)$ as the drop in $I_{\A:\S}$ after $\Me$ (see Fig. \ref{fig:informationg})
\begin{equation} \label{eq:backaction}
    \D(\BA_\Me) \equiv I_{\A:\S}\left( \ket{\Phi^+}\bra{\Phi^+}_{\A\S} \right) - I_{\A:\S}\left( \left(\idc_\A\otimes \BA_\Me \right) [\ket{\Phi^+}\bra{\Phi^+}_{\A\S}]\right).
\end{equation}

To see that these quantities make sense, we examine the qualitative behaviour of  $I_{\A:\Ob}$ and $I_{\A:\S}$. 
First we note that once we fix $\dim(\Hil_\S)=\dim(\Hil_\A)=d$, the initial correlations are $I_{\A:\S}(\ket{\Phi^+}\bra{\Phi^+}_{\A\S})=2\log d$. Since $\Ob$ is initially uncorrelated with $\A$ and $\S$, we also have $I_{\A:\S\Ob}(\ket{\Phi^+}\bra{\Phi^+}_{\A\S} \otimes \chi_\Ob) = 2\log d$.  The measurement interaction $\Me$ is a local operation on $\S\Ob$, so $I_{\A:\S\Ob}(\Me[\ket{\Phi^+}\bra{\Phi^+}_{\A\S}]) \leq 2 \log d$. Ignoring $\Ob$ to obtain the back-action $\BA_\Me$ is a local operation, so $I_{\A:\S}(\BA_\Me[\ket{\Phi^+}\bra{\Phi^+}_{\A\S}]) \leq 2 \log d$. Ignoring $\S$ to obtain the result channel $\R_\Me$ is also local, so $I_{\A:\Ob}(\R_\Me[\ket{\Phi^+}\bra{\Phi^+}_{\A\S}]) \leq 2 \log d$. As a result,
\begin{equation} \label{eq:information_gain_bounds}
    0 \leq \In(\R_\Me), \ \D(\BA_\Me) \leq 2 \log d.
\end{equation}

\subsection{Examples}\label{sec:examples}

Let us consider the simple case where each subsystem is a qubit.  Both $\En$ and $\Ob$ are initially in the state $\ket{0}$ and the maximally entangled $\A\S$ state is $\ket{\Phi^+}=\frac{1}{\sqrt{2}}[\ket{00}+\ket{11}]$. In the von Neumann scheme, the final $\A\SEO$ state is $\frac{1}{\sqrt{2}}[\ket{0000}+\ket{1111}]$, which gives $\In(\R_\Me)=\D(\BA_\Me)= \log 2$. 

For a swap \quesurement{}, the final $\A\S\Ob$ state is $\frac{1}{\sqrt{2}}[\ket{000}+\ket{101}]$, so the reduced $\A\Ob$ state is maximally entangled and $\In(\R_\Me)= 2 \log 2$ (this can also be seen by noting that $\R_\Me$ is the $\S\rightarrow \Ob$ identity). The channel $\BA_\Me$ takes all system states to $\ket{0}_\S$ so  recovery is impossible and the disturbance is maximal: $\D(\BA_\Me)=2 \log 2$. 

The decohered swap \quesurement{} yields $\frac{1}{2} [\ket{000}\bra{000} + \ket{101}\bra{101}]$ as the final $\A\S\Ob$ state. The final $\A\Ob$ state is the classically correlated state $\frac{1}{2} [\ket{00}\bra{00} + \ket{11}\bra{11}]$, which has $\In(\R_\Me) = \log 2$. The final $\A \S$ state is uncorrelated $\frac{1}{2} [\ket{0}\bra{0} + \ket{1}\bra{1}] \otimes \ket{0}\bra{0}$ and has $\D(\BA_\Me) = 2 \log 2$.

These results generalize naturally to $d$ dimensional systems. The von Neumann scheme has $\In(\R_\Me) = \D(\BA_\Me) = \log d$, which we call the \emph{Heisenberg limit}. The swap \quesurement{} has $\In(\R_\Me) = \D(\BA_\Me) = 2 \log d$ and the decohered swap \quesurement{} has $\In(\R_\Me) = \log d$ and $\D(\BA_\Me) = 2 \log d$. These results are summarised in Table \ref{tab:measurements}.

The fact that the information gain of a swap \quesurement{} is $2 \log d$ warrants an explanation. Intuitively, a swap should provide at least \emph{some} information, even if it is not obvious how much. It should provide at least as much information as any von Neumann \quesurement{}, because once the system state falls into the memory of the agent's \sensor{}, she is free to post-process it with any von Neumann \quesurement{} she sees fit. Yet a naive analysis of the swap \quesurement{} suggests it provides no information. To wit, the swap \quesurement{} maps any system state $\ket{\psi}$ to a separable state $\ket{0 \psi}$. The system and observer are completely uncorrelated, no matter the initial system state. Any attempt to quantify information gain based on correlations of observer and system observables will fail to match the intuition that a swap offers non-zero information. It is a triumph of the \quesurement{} formalism that it succeeds in assigning the swap non-zero information. In fact, it assigns the swap the highest possible value of $2 \log d$ for a given system dimension $d$.

\subsection{Information-disturbance relation}
\label{sec:information_disturbance_relation}

The examples above suggest a relationship between the information gain $\In(\R_\Me)$ and disturbance $\D(\BA_\Me)$ of a measurement interaction $\Me$. In the first two examples (von Neumann and swap) we had $\In(\R_\Me) = \D(\BA_\Me)$ and in the last example (decohered swap) we had $\D(\BA_\Me) > \In(\R_\Me)$. In all three examples, the observer disturbed at least as much information as she gained. In the following theorem, we prove that $0 \leq \D(\BA_\Me) - \In(\R_\Me)$. Furthermore, the examples suggest that when the \quesurement{} yields a pure state, the disturbance equals the information gain. We show that this is true in general by proving that the difference between the disturbance and information gain is bounded from above by twice the joint entropy of the $\ASO$ system.

\begin{theorem}\label{theo:information_disturbance}
Consider a measurement interaction $\Me : \S \rightarrow \S\Ob$ of a system $\S$ with a Hilbert space of finite dimension $d$. 
Let $\ket{\Phi^+}_{\A\S}$ be the state of that system maximally entangled with an ancilla $\A$. Let $\rho_{\A \S\Ob} = \left(\idc_\A \otimes \Me \right) [\ket{\Phi^+}\bra{\Phi^+}_{\A\S}]$ be the state after the interaction. Then the information gain
$\In(\R_\Me) = I_{\A:\Ob}\left( \rho_{\A \Ob} \right)$ 
and disturbance
$\D(\BA_\Me) = 2 \log d - I_{\A:\S}\left( \rho_{\A \S} \right)$ satisfy
\begin{equation}\label{eq:information_disturbance}
    0 \leq \D(\BA_\Me) - \In(\R_\Me) \leq 2 S(\rho_{\ASO})
\end{equation}

\end{theorem}

\begin{proof}

We start by proving $0 \leq \D(\BA_\Me) - \In(\R_\Me)$. Our proof is based on the strong subadditivity of quantum information, which can be expressed as the following inequality valid for any tripartite state: $\tau_{\mathcal{X}\mathcal{Y}\mathcal{Z}}$: \cite[Eq. 11.114]{NielsenandChuang}
\begin{equation}
    I_{\mathcal{X} : \mathcal{Y}}(\tau_{\mathcal{X}\mathcal{Y}}) + I_{\mathcal{X} : \mathcal{Z}}(\tau_{\mathcal{X}\mathcal{Z}}) \leq 2 S (\tau_{\mathcal{X}}).
\end{equation}
Applying this inequality to $\rho_{\ASO}$ shows that
\begin{equation}
     I_{\A : \Ob}(\rho_{\A\Ob}) \leq 2 S_{\A}(\rho_{\A}) - I_{\A : \S}(\rho_{\A\S}).
\end{equation}
The interaction $\Me$ does not act on the ancilla, so the $\A$ subsystem remains in its initial state, which was half of a maximally entangled state. Consequently, $S(\rho_\A) = \log d$. Substituting this value, along with the definitions for disturbance and information gain reveals $\In(\R_\Me) \leq \D(\BA_\Me)$.

Next we prove $\D(\BA_\Me) - \In(\R_\Me) \leq 2 S(\rho_{\ASO})$. The proof relies on a corollary of strong subadditivity known as the Araki-Lieb triangle inequality\cite{araki1970entropy}, which states that any bipartite state $\tau_{\mathcal{X}\mathcal{Y}}$ obeys the following inequality:
\begin{equation}
    |S(\tau_\mathcal{X}) - S(\tau_\mathcal{Y})| \leq S(\tau_{\mathcal{X}\mathcal{Y}}).
\end{equation}
We also reuse the fact that $S(\rho_\A) = \log d$.
\begin{align}
    \D(\BA_\Me) - \In(\R_\Me) 
    & = 2 \log d - I_{\A:\S}\left( \rho_{\A \S} \right) - I_{\A:\Ob}\left( \rho_{\A \Ob} \right ) \\
    & = 2 S(\rho_\A) - [ S(\rho_A) + S(\rho_\S) - S(\rho_{\A \S}) ] - [ S(\rho_A) + S(\rho_\Ob) - S(\rho_{\A \Ob}) ] \\
    & = [S(\rho_{\A \S}) - S(\rho_\Ob)] + [S(\rho_{\A \Ob}) - S(\rho_\S)] \\
    & \leq 2 S(\rho_{\ASO})
\end{align}
The last line follows from applying the triangle inequality to obtain $S(\rho_{\A \S}) - S(\rho_\Ob) \leq S(\rho_{\ASO})$ and $S(\rho_{\A \Ob}) - S(\rho_\S) \leq S(\rho_{\ASO})$.

\end{proof}

The upper bound in Eq. \ref{eq:information_disturbance} shows that a sufficient condition for a \quesurement{} to inform as much as it disturbs is for it to map pure state to pure states, and thus output an $\ASO$ state with zero entropy. However, this sufficient condition is not a necessary condition. The von Neumann measurement considered in Sec. \ref{sec:examples} produced the $\A\SEO$ state $\frac{1}{\sqrt{2}}[\ket{0000}+\ket{1111}]$. Although the reduced $\ASO$ state is mixed, the \quesurement{} has equal disturbance and information gain. A complete classification of the necessary and sufficient conditions for a \quesurement{} to inform as much as it disturbs would essentially give a recipe for ideal, minimally disturbing measurements, but it is beyond the scope of this work.

\section{Agency in the quantum world} \label{sec:agency}

In the previous sections we developed a framework to study measurements involving agents equipped with quantum \sensors{}. Within this framework we extended the classically motivated concepts of observation and uncertainty into the quantum regime without trying to force a connection to our own experience.  While our experience of the world seems to be classical at least in the sense that our observations can be described using POVMs,  there seems to be no law of nature that can  rule out the more general type of observation described in Sec. \ref{sec:quantumObs}  above  (It should however be emphasised that a wide body of work has been devoted to finding such laws; e.g. the  quantum Darwinism \cite{Brand_o_2015,Ollivier_2004,Zurek_2009} and spontaneous collapse \cite{Bassi_2013} programs). If we believe quantum theory applies at all levels, then quantum agents can exist in principle, for example in the form of  sufficiently advanced fault tolerant quantum computers with peripheral quantum sensors (see Fig. 
\ref{fig:networks}).  In this section we discuss quantum agency through well known results developed in the context of traditional measurements and explore possible directions for research in quantum computing.

\subsection{Simultaneous  observations} \label{sec:simultaneous}
\begin{figure}
    \centering
    \includegraphics[width=\textwidth]{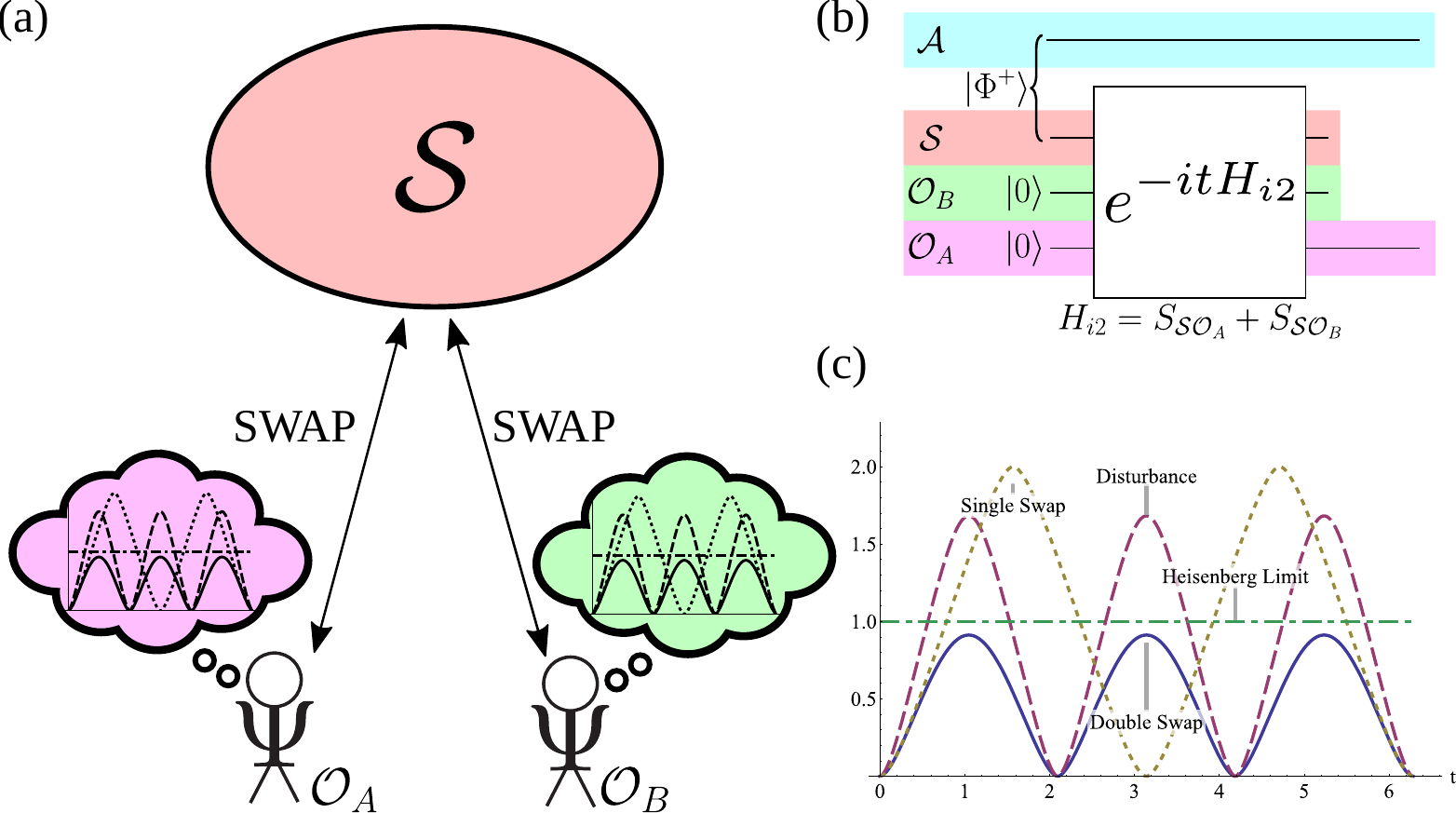} 
    \caption{{\bf Simultaneous observations using a swap interaction} (a) Two observers $\Ob_A$ and $\Ob_B$ simultaneously couple to the same system using the swap Hamiltonian. (b) A circuit diagram for the \quesurement{} with an ancilla $\A$ added to account for disturbance and information gain. (c) Informational quantities in bits as a function of time. The solid blue curve denotes the information gained by each observer individually during the double swap. Dashed purple shows the disturbance to the original system. It also equals the information gained by both observers when they cooperate and act as a single, joint observer. Olive dots denote the single swap information gain and disturbance, which describes what would have happened if only one observer attempted to sense the system. Green dashed dots show the Heisenberg limit for reference, which corresponds to the information gain and disturbance of an ideal von Neumann measurement. At the cost of extra disturbance compared to a von Neumann measurement, swap observers can nearly achieve the Heisenberg limit without cooperation and significantly exceed it with cooperation.}
    \label{fig:simultaneous}
\end{figure}

We now move to a scenario involving two observers $\Ob_A$ and $\Ob_B$ who try to sense the same system simultaneously  (see Fig. \ref{fig:simultaneous}). In the case of a von Neumann \quesurement{} of the same observable, this is a well studied scenario \cite{Zurek_2009}. The two \quesurements{} commute and the outcomes are the same as those arising from the scenario where the observers perform the \quesurement{} one after the other,  or a situation where the two observers monitor the same \sensor{}. In all of these cases the observers do not influence each other's \quesurement{} and all can reach the Heisenberg limit and moreover have correlated results.

What happens if two observers attempt to sense the same system simultaneously, both using the same swap \quesurement{}? The SWAP operator $S_{\S\Ob_A}$ that swaps system states with memory states in $\Ob_A$'s \sensor{} does not commute with the otherwise identical SWAP operator $S_{\S\Ob_B}$ for $\Ob_B$. As a result, the two observers' \quesurements{} will influence each other. We model the simultaneous swap using the interaction Hamiltonian $H_{i2}=S_{\S\Ob_A}+S_{\S \Ob_B}$. The system and observers are initially in the product state $\ket{\psi}_\S\ket{0}_{\Ob_A}\ket{0}_{\Ob_B}$. The Hamiltonian $H_{i2}$ is applied for a time $t$, leading to the state $ e^{-itH_{i2}}\ket{\psi}_\S\ket{0}_{\Ob_A}\ket{0}_{\Ob_B}$.  
At every moment in time, we associate a \quesurement{} with each observer, $\Ob_A$ and $\Ob_B$, denoted $\Me_A$ and $\Me_B$ respectively. 
Both of these \quesurements{} refer to the same dynamical process, save for an exchange in the notion of observer and environment.  
Since the situation is symmetric, we will only study it from observer $\Ob_A$'s perspective. 

We explicitly calculate the time-evolved state in \ref{eq:twoswappostmeasurement}. From Eq. \eqref{eq:twoswappostmeasurement}, we compute the information gain and disturbance and plot it for $d=2$ in Fig. \ref{fig:simultaneous} c. For comparison with this ``double swap'' scenario, we plot the information gain and disturbance for a ``single swap'' \quesurement{} in which only $\Ob_A$ observes the system and the interaction Hamiltonian is $H_{i1}=S_{\S\Ob_A}$. Initially, each observer gains nearly as much information from the double swap as the single swap. By the time each observer gains $0.5$ bits of information (half the Heisenberg Limit), the presence of the other observer is felt and the double swap information begins to lag behind the single swap. While the single swap reaches the maximal information gain and disturbance of $2$ bits at time $t = \pi / 2$, the double swap caps out sooner at $t = \pi / 3$ with an information gain of $0.91$ bits and a disturbance of $1.68$ bits. If each observer had instead performed the same von Neumann measurement on the system, they would have each gained $1$ bit of information, slightly more than the double swap. They also would have only disturbed the system by $1$ bit, significantly less than the double swap.

The double swap seems worse in terms of both information gain and disturbance than a von Neumann measurement, but there is more to the story. So far, we have only considered how much information $\Ob_A$ and $\Ob_B$ gain individually. However, if they cooperate we can treat them as a single joint observer and ask how much information $\Ob = \Ob_A \Ob_B$ gains. For this joint observer, the interaction with the system is unitary, so their information gain equals their disturbance and maximizes to $1.68$ bits. This information gain significantly exceeds the $1$ bit of information that one, two, or indeed any number of cooperative observers using the same von Neumann measurement can extract. Even though each von Neumann observer individually gains $1$ bit from their measurement, it is all the \emph{same}, redundant bit. At the cost of extra disturbance, two observers using a swap \quesurement{} can gain almost as much information individually as a von Neumann measurement while gaining significantly more information as a single, cooperative unit.

\subsection{Quantum computers as agents} \label{sec:qcomputers}

Work on mechanized observers who are part of the quantum system goes back at least as far as Everett \cite{Everett57} who imagined quantum automata observing the system in a generic way, but focused his attention on von Neumann type measurements and a classical experience where results are classical labels. Later, Albert \cite{Albert1983,Albert1987} showed that a quantum automaton  with access to its own memory registers could perform measurements whose (classically interpreted) results seem paradoxical. These works inspired much of the early theoretical work on quantum computing, in particular Deutsch's pioneering work on universal quantum Turing machines \cite{DeutschUniversal}.  Most  subsequent research however, regards quantum computers as devices to be used by classical agents. \footnote{Note that the possibility of having quantum computers as observers is sometimes mentioned in the context of quantum foundations \cite{Bong2019}.}

\begin{figure}
    \centering
    \includegraphics[width=\textwidth]{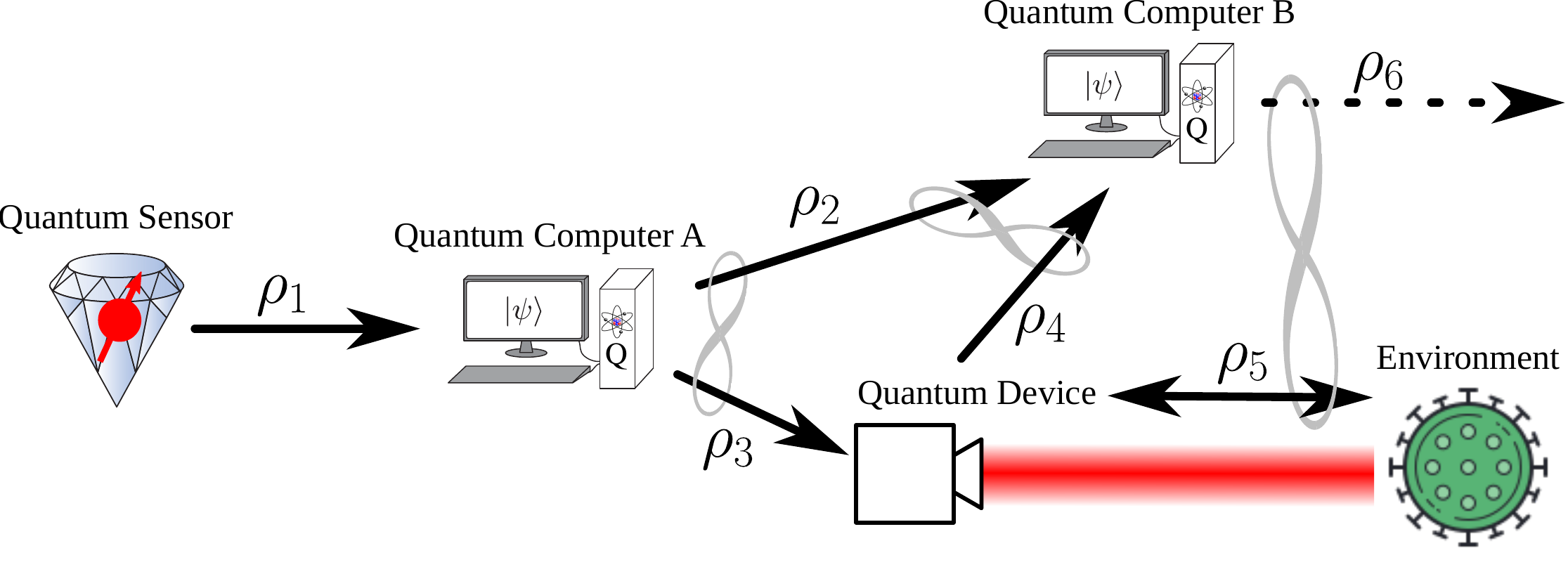}
    \caption{A quantum network where the different agents (the quantum computers) communicate via quantum communication channels and interact with the world via quantum peripherals. If we ascribe agency to sufficiently advanced quantum computers we expect that most of their interactions with each other and the rest of the world will be fully coherent. }
    \label{fig:networks}
\end{figure}

The usual quantum information processing paradigm involves classical inputs and outputs. In a sampling problem for example, the input is a classical description of some quantum circuit ending with a sequence of binary measurements, and the output is  a string of bits that represents a sample from the probability distribution for the measurement outcomes.  This picture is certainly realistic, but it can be modified in a quantum internet  scenario where  quantum computers are  directly connected to quantum sensors that feed  a quantum state as the input for a computation. These computers could  interact with other quantum  computers through  quantum interconnects, so that they only share quantum information. If perfect quantum communication networks were available, we would expect most of the communication between quantum computers to involve no classical results (see Fig. \ref{fig:networks}). Quantum states generated by one computer could then be used to make decisions about which circuit to encode on other computers. While, most circuits are realized presently with classical control parameters, quantum controls have been used to run superpositions of circuits \cite{Chiribella_2019, Rubino2021, pang2023experimental}. Quantum computers communicating in this way would have entangled circuits and classical results would only appear when these computers interact with us. \footnote{ and perhaps never if certain science fiction scenarios turn out to be correct.} Such communication would allow many small quantum computing nodes to be joined together into a single distributed quantum computer.

Work on quantum computing theory tends to be based on the assumption that quantum theory can be applied as-is to arbitrarily large systems, yet the vast majority of work takes an anthropocentric view: classical input, classical output. Notably this approach even extends to problems in the QMA computational complexity class \cite{aharonov2002quantum}  which involve a proof encoded in a quantum state that is sent to a verifier who outputs a classical ``accept" or ``reject" with an appropriate probability. Exceptions include delegated quantum computing \cite{tham2020experimental,Fitzsimons_2017} and compression protocols \cite{rozema2014quantum,Pepper_2019,Yang_2016} which are usually framed  as intermediate stages in a larger classical-in classical-out task. To the best of our knowledge, none of these have been studied as computational complexity classes. For example, one might study a type of decision problem where the language is defined in Hilbert space and the verifier's output is a quantum state.

There are a number of difficulties in extending ideas in computational complexity to the fully quantum regime. One technical issue involves  dealing with imperfections and handling non-orthogonal output states. Classically it is customary to expect that the right result would appear with a high probability (say 2/3) and that the results can be distinguished from each other. In the quantum case, one could try to deal with imperfections by asking that the real output state be close to some ideal output state. The metric used for the distance should make sense in the context of the problem. The usual distance measures used in quantum mechanics are interpreted in terms of probabilities for measurement outcomes, however as we showed here (Sec. \ref{Sec:uncertainty}), these can still make sense in a theory without probabilities.  The same issue appears when dealing with non-orthogonal output states. There seems to be no inherent reason to claim that orthogonal states are more distinguishable to a quantum observer than non-orthogonal ones. More precisely: without the Born rule, there is no obvious notion of distinguishability which can be applied to orthogonal states.  It might however be reasonable to expect that since unitary operators preserve the inner product, it has a special meaning. In fact this is probably unavoidable once we assign meaning to reduced density operators, or start thinking about broadcasting information to multiple observers \cite{Deutsch_2015}.

\section{conclusions} \label{sec:conclusions} 

The orthodox approach to measurement in quantum mechanics is based on an assumption that the observer is external to the quantum system and can observe the quantum world through \emph{observables}. This  approach  works incredibly well to describe all known quantum phenomena and is unlikely to  fail as long as observers (such as ourselves) have classical memories. Quantum memories do not seem to be a naturally occurring phenomenon, but much of the effort in quantum information processing is aimed at increasing coherence times and it is increasingly likely that engineered quantum memories will become commonplace in the not-too-distant future.

If we are to believe that the current technological trajectory will continue, we should entertain the idea of attributing agency to  sufficiently advanced quantum computers.  As we have shown in this work, such quantum agents would have access to a broader range of actions than those available to us. In particular, they would use \quesurements{} rather than conform to the narrow definition of measurement which has been adopted in quantum theory. We have explored these ideas and presented the swap \quesurement{}  (Sec. \ref{sec:swap} and Fig. \ref{fig:examples} b) as an extreme case of  quantum observation. Our ideas have already inspired an upcoming experiment \cite{pang2024agency} which empirically measures the information flow of several \quesurements{}.

We began with a generalized definition of measurement (Def. \ref{def:meas}) whose deterministic result is a quantum state recorded in the quantum memory of an observer's \sensor{}. To handle this more general formalism (which includes the von Neumann scheme as a special case) we modified the basic mathematical tools of quantum measurement theory and replaced the POVM with the result channel $\R_\Me$, which takes system states to states of the memory in the observer's \sensor{} (see Fig. \ref{fig:quantummeasurement}). 

Using these tools, we suggested a method for quantifying information gain and disturbance (Sec. \ref{sec:informationgain} and Fig. \ref{fig:informationg}). These definitions led to expected results for von Neumann measurements while faithfully describing the swap \quesurement{}'s exotic ability to provide complete information at the cost of maximal disturbance. We showed that the disturbance of a \quesurement{} always equals or exceeds the information gained, which places a fundamental constraint on the information processing power of quantum agents. We provided a sufficient condition for a \quesurement{} to be ideal in the sense that it informs as much as it disturbs, but left open the task of identifying conditions that are both sufficient and necessary. We used these quantities to analyze a thought experiment in which two agents both attempt to use a swap \quesurement{} on the same system at the same time. The thought experiment revealed that cooperation amongst quantum agents is essential for making the most of certain \quesurements{}.

Our results are the first steps in studying fully quantum agents, but there is still much work to be done. We have only studied how quantum agents observe their surroundings, but the processes by which quantum agents might decide and act using those observations is still ripe for exploration. While much work is being done on studying new mechanisms for decoherence-free interactions, there is very little work on protocols, algorithms, and computational paradigms that involve purely quantum agents. While our framework formally separates the observer from the observed, the physical mechanism by which a quantum agent gains a sense of self is still a mystery.

We hope that our work encourages others to consider scenarios that involve multiple quantum agents who can interact with each other in a purely coherent, quantum mechanical way, for example a world-wide network of intelligent quantum computers (see Fig. \ref{fig:networks}). As we work towards building sophisticated quantum machines and a quantum internet, we should give careful consideration to how these machines would \emph{think} and interact with the world around them. Learning from past oversights, we should not ignore the possibility of information being truly quantum mechanical.

\section{Acknowledgements}
We thank J. E. Sipe for useful discussions. This work was funded by grant number FQXi-RFP-1819 from the Foundational Questions Institute and Fetzer Franklin Fund, a donor advised fund of Silicon Valley Community Foundation. All authors were also supported by by  NSERC and the Fetzer Franklin Fund of the John E. Fetzer Memorial Trust.  A.M.S. is a fellow of CIFAR.

\appendix

\section{Traditional quantum measurements}

\subsection{Observables, outcomes and probabilities}

Measurements in quantum mechanics have traditionally been associated with `observables' which are mathematically represented as Hermitian operators. Each observable $A$  has unique set of eigenstates (eigenspaces if it is degenerate) and eigenvalues so that it can be written as $A=\sum_k a_k A_k$  where $a_j\ne a_k$ unless $j=k$ and $\{A_k\}$ are projectors onto orthogonal subspaces. The (real) eigenvalues $\{a_k\}$ are usually used to label the possible measurement results, while the projectors $\{A_k\}$ are used for calculating the probabilities for each result. For a system initially in the state $\rho$, the probability that a measurement of $A$ will yield the result $a_k$ is given by the Born rule $
p(a_k)=\tr(A_k\rho)$

 This formalism can be extended  by noting that the set of orthogonal projectors $\{A_k\}$ can be replaced by a set of positive operators  $\{E_k\}$ (not necessarily orthogonal) with $\sum_k  E_k = \openone$. Such a set of positive operators is called a positive operator valued measure (POVM) and its elements can  be plugged into the Born rule   to produce a probability distribution
  \begin{equation}\label{eq:Born}
     p(a_k)=\tr(E_k\rho)
 \end{equation} for a set of possible measurement results labeled $\{a_k\}$. For simplicity we will refer to Eq. \ref{eq:Born} as the Born rule.

 In many cases we are also interested in the back-action of the measurement on the measured system. The POVM does not provide sufficient information to predict a
 unique outgoing system state, although it can be used to identify a \emph{minimally disturbing} measurement  (as defined for example in \cite{Wisemanbook}). For a minimally disturbing measurement, a result $a_k$ with an associated POVM element $E_k$ implies the transformation $\rho\rightarrow \frac{\sqrt{E_k}\rho\sqrt{E_k}}{\sqrt{\tr(E_k\rho)}}$ on the measured system (see Fig. \ref{fig:vNmeasurement} a). In the special case of a measurement of an observable $A=\sum_k a_k A_k$, the minimally disturbing measurement is called a von Neumann measurement and the update rule is $\rho\rightarrow \frac{A_k\rho A_k}{\sqrt{\tr(A_k\rho)}}$. When $A_k$ is a rank-1 projector this rule has the simple form  $\rho\rightarrow A_k $ i.e., the state projects onto the eigenstate associated with $a_k$. The disturbance is considered minimal since the update rule leaves eigenstates of $A$ undisturbed.

 The above description of a measurement (Born rule and state update rule) is in most cases sufficient for making predictions about the outcomes of experiments where the precise details of the measurement procedure and the observer can be ignored. It does however imply a hard cut between the observer and the measured system. 

 \subsection{The von Neumann scheme}
 \label{sec:vonNeumannSCheme}
 
Von Neumann's approach can be used to model the measurement process associated with any POVM by treating a measurement apparatus as a quantum mechanical system initially in a state $\Upsilon$. The measurement begins with some interaction $U_\SM$ so that  $\rho \otimes \Upsilon \rightarrow U_\SM (\rho \otimes \Upsilon) U_\SM^\dagger$ after which the measurement result is encoded in the state of $\M$.  To  read out  the result, an observer would need to  measure an observable $\Pi_\M=\sum_k  a_k \Pi_k$  on $\M$, where $\Pi_k$ are orthogonal projectors and the labels $a_k$ are distinct.  The probability for a result $a_k$ would then be $P(a_k|\rho_\S) = \tr\left[(\openone \otimes \Pi_k) U_\SM (\rho \otimes \Upsilon) U_\SM^\dagger \right]$. This  equation can be written in the form of Born's rule  \eqref{eq:Born}  by identifying the  POVM element $E_k = \tr_\M\left [ U_\SM^\dagger (\openone \otimes \Pi_k) U_\SM (\openone \otimes \Upsilon)  \right ]$ so that $P(a_k|\rho_\S)=\tr \left[E_k \rho_\S \right ]$. The procedure also gives the post-measurement  state of the observed system given a result $a_k$ as $\frac{\tr_\M\left[(\openone \otimes \Pi_k) U_\SM (\rho \otimes \Upsilon) U_\SM^\dagger \right]}{P(a_k|\rho_\S)}$\cite{Ozawa1984}.   The term \emph{von Neumann scheme} is often used for the special case where this procedure is applied to the inner workings of a  von Neumann measurement. For this measurement $U_\SM$ is generated by a Hamiltonian in the form of Eq.  \eqref{eq:vonN} (see Fig. \ref{fig:vNmeasurement} above and Sec.  \ref{sec:vNexample} below for more details).

The scheme above is more detailed than the Born and state update rules, but it is not complete since it invokes an observation  of the measurement device (via  $\Pi_\M$)  with no details on how this measurement is constructed. It then begs the question {\it `how is the measurement device observed?'} to which we could give the same answer ad infinitum. At this point von Neumann invoked the external observer (which he previously justified).  It is however possible to treat the result  quantum mechanically - i.e. as a state $\ket{a_k}$ rather than a classical label `$a_k$' -  so that no external observers are required. This fully coherent approach comes with interpretational issues, but it can be argued that these are no more problematic than the alternative. As we will show in Sec. \ref{sec:quantumObs} the possibility of encoding the measurement result in a quantum state allows a more general definition of measurement where some observers can perceive the world in a way which cannot be modeled in the language of POVMs.

\subsection{Side remark: Locality and the Born rule}

Quantum mechanics with external (classical) observers is fairly well defined at least from an operational perspective where certain systems (e.g. humans, cameras, etc.) are postulated to be classical and external. In such an approach one generally accepts the Born rule and a certain version of the collapse postulate as part of the quantum-to-classical transition.  This in turn allows us to  use some tools which we usually take for granted, especially those that have been adopted from statistical mechanics and information theory. One which is of particular significance here is the reduced state. Consider a system which is composed of two distinct subsystems: $\S$ and $\M$ such that its description is a state  $\ket{\psi}_\SM$  in the tensor product Hilbert space $\Hil_\SM=\Hil_\S\otimes\Hil_\M$. The reduced state of the system is $\rho_\S=\tr_\M[ \ket{\psi}\Bra{\psi}]$, where $\tr_\M$ is the partial trace over $\Hil_\M$. This reduced state contains all the information  necessary for calculating probabilities for the outcomes of measurements via the Born rule \eqref{eq:Born}.  From the operational perspective  of the external observer, this reduced state contains all there is to know about the system. %With this formal notion of the local state we will generally treat states as positive-semidefinite operators on the relevant Hilbert spaces. 
The notion of a reduced state can also be extended to dynamics. Consider an initial product state $\rho_\S\otimes\tau_\M$ on $\Hil_\SM=\Hil_\S\otimes\Hil_\M$ and some Schr\"odinger evolution described by the unitary $U_\SM$.  The local (reduced) dynamics is described by the completely positive trace preserving  map  $\mathfrak{C} (\rho_\S)=\tr_\M [U_\SM\rho_\S\otimes\tau_\M U^\dagger_\SM]$ (also called the \emph{quantum channel}).

Quantum theory with quantum observers provides a more difficult situation than a theory with external observers, and requires a more careful treatment of fundamental postulates and their interpretation. Our approach here is to take the `standard' Schr\"odinger-picture quantum theory (without collapse) at face value, as is often done in many-worlds interpretations\footnote{see  Deutsch and Hayden \cite{Deutsch_2000} for a different (Heisenberg based) approach to many worlds.} \cite{sep-qm-manyworlds,McQueen2019}. In particular, we assume that the reduced density operators are  valid complete descriptions of the local states, and that reduced dynamics provide a complete description of the local dynamics. We note that this is not necessarily the situation in hidden variable theories such as those following de Broglie and Bohm's pilot wave \cite{Bohm1952},  and that even in the many worlds interpretation the justification for assigning meaning to reduced density operators requires a complicated argument and non-trivial assumptions \cite{Sebens2018}. 

\section{Photodetection}
\label{sec:photodetection}

The usual treatment of photodetection \cite{GlauberDetectors,GerryKnight} begins with an atomic system interacting with an electric field through an interaction Hamiltonian of the form $H_i= \vec{d}\cdot \vec{E}$, where $\vec{d}$ is the dipole moment of the atom and $\vec{E}$ is the electric field. This Hamiltonian  has  the von Neumann form \eqref{eq:vonN} with the atom acting as the measurement apparatus and the field $\vec{E}$ as the measured operator, and so we might be tempted to say that this is a von Neumann measurement of the field. However, free evolution cannot  be neglected  at optical frequencies,  and the free evolution term in the Hamiltonian does not commute with the interaction term. The  dynamics are usually  treated by moving to a rotating frame and making the rotating wave approximation to get rid of terms that oscillate rapidly (see  \cite[Sec. 4.3 ]{GerryKnight} ).

We treat the detector as a two-level atom and assume that the atoms are initially in the ground state and that the excited state is an ion and a free photoelectron. Each absorbed photon leads to the release of a photoelectron (see \cite[Sec. 5.2]{GerryKnight} for a detailed derivation) which induces an irreversible amplification sequence whose details depend on the specifics of the detector.  Ideally the detector would perform a perfect  SWAP followed by dephasing so that the incoming photon number would be perfectly correlated with the amplified output signal; however, practical issues mean that photon number sensitivity is usually far from perfect. 

\section{Simultaneous Swap}

We calculate the state evolution and information exchange of the simultaneous swap \quesurement{} described in Sec. \ref{sec:simultaneous}. There are two relevant (unnormalized) eigenstates of $H_{i2}$:
\begin{align}\label{eq:eigenstates}
    \ket{e_{2}} &= \ket{\psi00} + \ket{0\psi0} + \ket{00\psi} ,\\\nonumber 
    \ket{e_{-1}} &= 2\ket{\psi00} - \ket{0\psi0} - \ket{00\psi}.
\end{align}
They have respective eigenvalues $2$ and $-1$. These allow us to expand the initial state $\ket{\psi 00}$ in the eigenbasis of $H_{i2}$.
\begin{equation}
    \ket{\psi 00} = \frac{\ket{e_2} + \ket{e_{-1}}}{3}
\end{equation}
The evolved state is
\begin{align} \label{eq:doublemeasurement}
    e^{-itH_{i2}}\ket{\psi00} 
    & = \frac{e^{-2it} \ket{e_2} + e^{it} \ket{e_{-1}}}{3} \\\nonumber
    & = \frac{e^{-2it} + 2 e^{it}}{3} \ket{\psi00} + 
    \frac{e^{-2it} - e^{it}}{3} \left [ \ket{0\psi0} + \ket{00\psi} \right ].
\end{align}
When $t$ is an integer multiple of $\frac{2}{3}\pi$, the state evolves back to its original uncorrelated state times a phase factor. For these $t$ values, the interaction is not a \quesurement{} by Def. \ref{def:meas}.

To calculate information gain and disturbance for the $d=2$ case, we add a third agent $\A$ initially entangled with $\S$, so that the initial $\A\S\Ob_A\Ob_B$ state is 
\begin{equation}
    \ket{\Psi}=\frac{\ket{0000}+\ket{1100}}{\sqrt{2}}.
\end{equation}
The state evolves to
\begin{equation}\label{eq:twoswappostmeasurement}
e^{-itH_i}\ket{\Psi} = \frac{1}{\sqrt{2}} \left [ e^{-2it} \ket{0000} + \frac{e^{-2it} + 2 e^{it}}{3} \ket{1100} + \frac{e^{-2it} - e^{it}}{3} \left [ \ket{1010} + \ket{1001} \right ] \right ].
\end{equation}
The information gain is the mutual information of the reduced $\A \Ob_A$ state,
\begin{equation}
    \rho_{\A \Ob_A} = \frac{1}{2} \left [\ket{00} + \frac{1 - e^{3it}}{3} \ket{11} \right ] \left [\bra{00} + \frac{1 - e^{-3it}}{3} \bra{11} \right ] + \frac{7 + 2 \cos(3 t)}{18} \ket{10}\bra{10}.
\end{equation}
The disturbance is $2$ minus the mutual information of the reduced $\A \S$ state,
\begin{equation}
    \rho_{\A \S} = \frac{1}{2} \left [\ket{00} + \frac{1 + 2 e^{3it}}{3} \ket{11} \right ] \left [\bra{00} + \frac{1 + 2 e^{-3it}}{3} \bra{11} \right ] + \frac{2 - 2 \cos(3 t)}{9} \ket{10}\bra{10}.
\end{equation}

\bibliographystyle{unsrt}
\bibliography{cnvs}

\end{document}